\pdfoutput=1
\documentclass[%
 reprint,
superscriptaddress,
 amsmath,amssymb,
 aps,
]{revtex4-2}

\DeclareUnicodeCharacter{2212}{\ensuremath{-}}

\usepackage{graphicx}
\usepackage{dcolumn}
\usepackage{bm}
\usepackage{color}
\UseRawInputEncoding

\begin{document}


\title{Colossal Magnetoresistance in Twisted Intertwined Graphene Spirals}

\author{Yiwen Zhang}
\homepage{These authors contributed equally to this work.}
\affiliation{School of Physical Science and Technology, ShanghaiTech University, Shanghai 201210, China}
\affiliation{ShanghaiTech Laboratory for Topological Physics, ShanghaiTech University, Shanghai 201210, China}
\author{Bo Xie}
\homepage{These authors contributed equally to this work.}
\affiliation{School of Physical Science and Technology, ShanghaiTech University, Shanghai 201210, China}
\affiliation{ShanghaiTech Laboratory for Topological Physics, ShanghaiTech University, Shanghai 201210, China}

\author{Yue Yang}
\homepage{These authors contributed equally to this work.}
\affiliation{School of Physical Science and Technology, ShanghaiTech University, Shanghai 201210, China}

\author{Yueshen Wu}
\homepage{These authors contributed equally to this work.}
\affiliation{School of Physical Science and Technology, ShanghaiTech University, Shanghai 201210, China}
\affiliation{ShanghaiTech Laboratory for Topological Physics, ShanghaiTech University, Shanghai 201210, China}

\author{Xin Lu}
\affiliation{School of Physical Science and Technology, ShanghaiTech University, Shanghai 201210, China}

\author{Yuxiong Hu}
\affiliation{School of Physical Science and Technology, ShanghaiTech University, Shanghai 201210, China}

\author{Yifan Ding}
\author{Jiadian He}
\author{Peng Dong}
\author{Jinghui Wang}
\author{Xiang Zhou}
\affiliation{School of Physical Science and Technology, ShanghaiTech University, Shanghai 201210, China}
\affiliation{ShanghaiTech Laboratory for Topological Physics, ShanghaiTech University, Shanghai 201210, China}

\author{Peng Dong}
\author{Yiwen Zhang}
\author{Yueshen Wu}
\author{Xiang Zhou}
\author{Jinghui Wang}

\affiliation{School of Physical Science and Technology, ShanghaiTech University, Shanghai 201210, China}
\affiliation{ShanghaiTech Laboratory for Topological Physics, ShanghaiTech University, Shanghai 201210, China}

\author{Jianpeng Liu}
\email{liujp@shanghaitech.edu.cn}
\affiliation{School of Physical Science and Technology, ShanghaiTech University, Shanghai 201210, China}
\affiliation{ShanghaiTech Laboratory for Topological Physics, ShanghaiTech University, Shanghai 201210, China}
\affiliation{Liaoning Academy of Materials, Shenyang 110167, China} 

\author{Zhu-Jun Wang}
\email{wangzhj3@shanghaitech.edu.cn}
\affiliation{School of Physical Science and Technology, ShanghaiTech University, Shanghai 201210, China}

\author{Jun Li}
\email{lijun3@shanghaitech.edu.cn}
\affiliation{School of Physical Science and Technology, ShanghaiTech University, Shanghai 201210, China}
\affiliation{ShanghaiTech Laboratory for Topological Physics, ShanghaiTech University, Shanghai 201210, China}




\date{\today}

\begin{abstract}
Colossal magnetoresistance (CMR) is highly applicable in spintronic devices such as magnetic sensors, magnetic memory, and hard drives. Typically, CMR is found in Weyl semimetals characterized by perfect electron-hole symmetry or exceptionally high electric conductivity and mobility. Our study explores this phenomenon in a recently developed graphene moir$\acute{e}$ system, which demonstrates CMR owing to its topological structure and high-quality crystal formation.  We specifically investigate the electronic properties of three-dimensional (3D) intertwined twisted graphene spirals (TGS), manipulating the screw dislocation axis to achieve a rotation angle of 7.3$^{\circ}$. Notably, at 14 T and 2 K, the magnetoresistance of these structures reached 1.7$\times$10$^7$\%, accompanied by an unexpected metal-to-insulator transition as the temperature increased. This transition becomes noticeable when the magnetic field exceeds a minimal threshold of approximately 0.1 T. These observations suggest the existence of complex, correlated states within the partially filled three-dimensional Landau levels of the 3D TGS system. Our findings open up new possibilities for achieving CMR by engineering the topological structure of 2D layered moir$\acute{e}$ systems.

\end{abstract}

\maketitle


\section{INTRODUCTION}

Giant magnetoresistance (GMR), typically below 100 percent, is common in thin-film metals and manganese-based perovskite, playing a crucial role in spintronic devices\cite{RN220,RN222}. The demand for higher performance has led to a focus on colossal magnetoresistance (CMR), characterized by a substantial increase in resistivity under magnetic fields, without saturation even at very high fields and low temperatures\cite{RN224,RN43,RN225,RN232,RN226,RN227,RN228}. This CMR, fundamentally distinct from GMR, is often observed in semimetals with topological properties and balanced electron-hole carriers as collected in Fig. 1b\cite{RN224}.

Notably, van der Waals semimetals like WTe$_2$\cite{RN43}, MoTe$_2$\cite{RN225,RN232}, and NbSb$_2$\cite{RN226} display CMR (1$\times$10$^5$\% at 2 K). Angle-resolved photoemission spectroscopy (ARPES) studies in these materials reveal electron and hole Fermi surfaces of equal size\cite{RN227}, indicating a resonant compensated semimetal nature. Similarly, the 3D transition metal diphosphides, WP$_2$ and MoP$_2$, showcase promising CMR due to closely neighboring Weyl points. Other semimetals like LaSb\cite{RN228}, TaAs\cite{RN229,RN231}, NbP\cite{RN23}, Cd$_3$As\cite{RN103}, and PtSn$_4$\cite{RN230} also exhibit CMR, mainly due to nearly perfect electron-hole compensation or high conductivity and mobility. 

Graphite, a fundamental semimetal, has shown impressive CMR progress, attributed to its topological defects such as stacking faults, and sample quality\cite{RN61}. In twisted graphene/boron–nitride (BN) heterostructure\cite{RN25}, magnetoresistance (MR) values reach 880\% at 400 K (9 T). A CMR of 110\% at 300 K (0.1 T) was reported in twisted BN/graphene/BN, linked to Dirac plasma in the moir$\acute{e}$ system\cite{RN34}. A record MR value of 1$\times$10$^7$\% (21 T) in micrometer-sized natural graphite suggests the importance of moir$\acute{e}$ patterns in CMR.
The 2018 discovery of superconductivity in magic-angle twisted bilayer graphene marked a milestone\cite{RN83,RN12}, making moir$\acute{e}$ superlattices a key focus in condensed matter physics\cite{RN51,RN68,RN72,RN69,RN54}. These layers exhibit unique phenomena due to the interplay between graphene’s Dirac wavefunctions and the interlayer moir$\acute{e}$ potential\cite{RN70,RN91,RN18,RN13}, leading to flat bands with nontrivial topology\cite{RN68,RN53,RN56,RN9,RN86,RN6,RN60,RN26,RN27}. Recently, a three-dimensional bulk moir$\acute{e}$ graphite superlattice, consisting of alternating twisted graphene layers, was synthesized\cite{RN11}. This structure, theoretically capable of 'magic momenta' and 3D Landau levels\cite{RN45}, may be a new platform for studying correlated and topological physics. 

Our study explores the transport properties of this new bulk moir$\acute{e}$ graphite, comprising hundreds and thousands of alternating twisted graphene multilayers, stacked along a spiral dislocation, as schematically shown in Fig. 1(a). The twisted graphene spirals (TGS) exhibit extremely high carrier mobility ($\mu_m$ = 3$\times$10$^6$ cm$^2$V$^{-1}$s$^{-1}$), forming Landau levels under low magnetic fields ($\sim$0.1 T). We observed a colossal magnetoresistance of 1.7$\times$10$^7$\% at 2 K under a 14 T magnetic field in this 3D TGS system.

\section{Result and Disscusion}
\subsection{Colossal magnetoresistance}
In Fig. 1b we present the temperature dependent MRs (at B= 7 T) of several typical high-MR materials including Weyl semimetals\cite{RN43,RN225,RN232,RN227,RN228,RN229,RN21}, Dirac semimetals\cite{RN226,RN103,RN230,RN22}, nodal-line semimetals\cite{RN24} (according to the degeneracy and momentum space distribution of the nodal points), NG and HOPG\cite{RN61,RN63}, graphene\cite{RN25,RN33}, and twisted spiral graphene presented in this work. Clearly twisted spiral graphene holds the record-high MR in this moderate magnetic-field regime. The MR at T = 2 K and B = 7 T is up to about 5$\times$10$^6$\%, which is a record high value among all MRs measured at moderate magnetic fields (\textless$\sim$10 T) reported in literatures. The value of MR in another sample reaches 1.7$\times$10$^7$\% with applied field B = 14 T at T = 2 K.  
For large MR, various potential mechanisms have been proposed, including electron-hole compensation, steep band, ultrahigh mobility, high residual resistivity ratio (RRR), topological fermions, etc. Mechanisms mentioned above can be analyzed through quantum oscillation and Hall effect measurement. Furthermore, a large MR is correlated with other systems, such as 2D materials, which carriers are limited to the 2D plane and the electronic structure largely depended on the thickness of material.  However, the MR of the highly oriented pyrolytic graphite (HOPG) reveals MR of only about 1$\times$10$^1$\% \cite{RN61,RN63,RN33,RN71}, being several orders of magnitudes smaller than that of the bulk twisted GS. Nevertheless, an extremely high MR of about1$\times$10$^6$\% (2 K, 7 T) has also observed in the Sri Lankan natural graphite (NG), for which the existence of highly conducting 2D interfaces aligned parallel to the graphene planes are attributed to the origin\cite{RN61,RN63,RN33,RN71}.

Another system concentrated here is topological materials, which are characterized by symmetry protected band crossing at or near the Fermi level in the Brillouin zone. Particularly, the present record of the maximum MR was reported in type-II Weyl semimetal WP$_2$ (2$\times$10$^6$\%, 7 T, 2 K)\cite{RN21}.  Such spectacularly large MR in WP$_2$ were attributed to both the nontrivial wavefunction textures around Weyl nodes and the presence of compensating electron and hole pockets in the Fermi surface, where the former can suppress back scatterings and induce considerably large mobilities\cite{RN23,RN21,RN103,RN102}.As reported in nature graphite U11\cite{RN61}, the large MR in graphite shares different origin from Weyl semimetals. Due to the advanced growth method proliferates the scope of graphene moir$\acute{e}$ superlattices\cite{RN11}, bulk graphite comprises alternatingly twisted graphene multilayers, which are sequentially stacked along a spiral dislocation. While the topological structure of GS still deserves discussion and the structure of the GS will be introduced at the following section.

\begin{figure*}[!htbp]
\includegraphics[width=0.8\textwidth,clip]{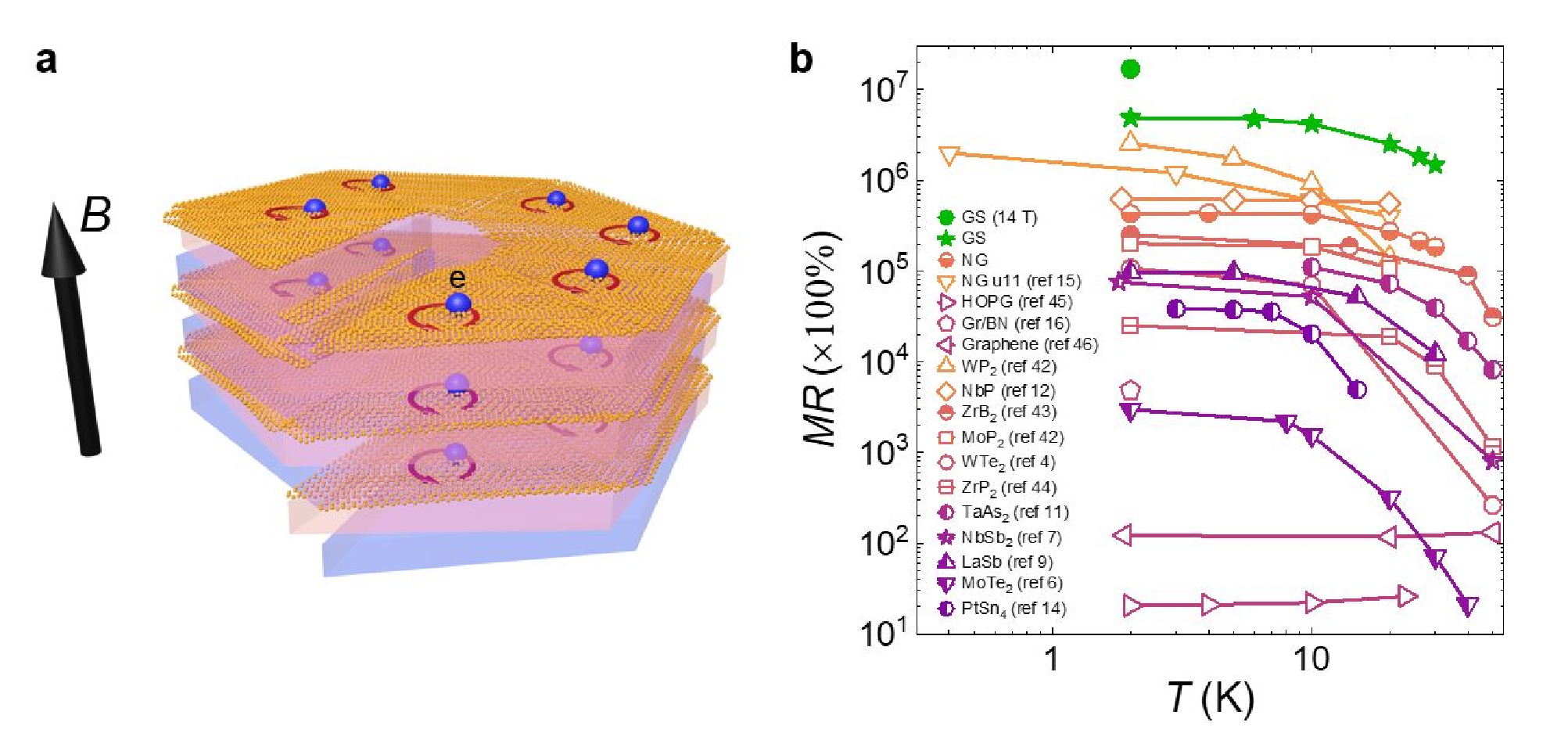}
\centering
\caption{(color online).  Colossal magnetoresistance in various semimetals at an applied field of 7 T. (a) Illustration of charge transport in GS. (b) The temperature dependence of larger MR for GS and some other typical materials, including the topological semimetals WP$_2$\cite{RN21}, NbP\cite{RN23}, MoP$_2$\cite{RN21}, WTe$_2$\cite{RN43}, ZrP$_2$\cite{RN24}, ZrB$_2$\cite{RN22}, TaAs$_2$\cite{RN231}, NbSb$_2$\cite{RN226}, LaSb\cite{RN228}, MoTe$_2$\cite{RN232}, and PtSn$_4$\cite{RN230}, highly oriented pyrolytic graphite (HOPG)\cite{RN63}, natural graphite (NG)\cite{RN61}, graphene\cite{RN33}, and graphene/BN moir$\acute{e}$ lattice\cite{RN25}. 
}
\end{figure*}

\subsection{Structure of the twisted graphene spira}

In the scanning electron microscopy (SEM) imaging, we observe stepwise variations in contrast, allowing us to discern up to ten individual graphene layers \cite{RN66}. Given that each graphene spiral (GS) in crystal is comprised of several dozen turns, SEM thus is not adept at distinguishing the contours of GS with up to ten layers. To circumvent this limitation, we resort to confocal laser scanning microscopy (CLSM), which possesses the ability to simultaneously generate intensity and topographical images. This facilitates the measurement of the in-plane dimensions and the out-of-plane stacking sequence of the as-grown GSs \cite{RN75}. Through CLSM, we confirm that the chemical vapor deposition (CVD) approach is proficient at producing GSs with dimensions spanning hundreds of micrometers, revealing interlayers characterized by an approximate twist angle of 7$^\circ$ (as showcased in Figure 2a and Fig. S8). Additionally, the adjacent zigzag edges of these spirals tilt at an angle of 7$^\circ$, a detail captured in the atomic force microscopy (AFM) imagery presented in Fig. S8. Such evidence intimates that the freshly synthesized GSs adopt an intertwined helical configuration, characterized by a twist of 7$^\circ$.

To further validate the presence of an intertwined helical structure with a homogeneous twist angle, we transferred the identical GS onto a transmission electron microscopy (TEM) grid for TEM diffraction measurements, following the transport measurements (which will be discussed later). The TEM observations, along with the corresponding selected area electron diffraction (SAED) patterns, reveal that the twist angle of the GS is 7.3$\pm$0.176$^\circ$ (Fig. 2b and Fig. S8). Significantly, the twist angle determined through SAED is consistent with the findings from CLSM and AFM observations in Fig. 2a and Fig. S8. More accurate twist angle measurements were made by diffraction mode of TEM, which allowing for the observation of superlattices formed by a pair of graphene lattices in the diffraction pattern, as depicted in Fig. 2b top inset. It is worth mentioning that electron diffraction of moir$\acute{e}$ patterns can also be observed, exhibiting the same symmetry as the original lattice.

The relative rotation of the GS leads to the formation of a periodic moir$\acute{e}$ superlattice within the helical graphene structure. The wavelength of this superlattice ($\lambda$) is determined by the twisted angle ($\theta$) present in the moir$\acute{e}$ superlattice, $\lambda$=$a$/(2sin($\theta$/2)) ($a$: the lattice constant of the monolayer graphene)\cite{RN59}. High-resolution TEM characterization reveals the presence of a moir$\acute{e}$ pattern with a twist angle of $\theta$ = 7.3$^\circ$, corresponding to a superlattice wavelength of approximately 1.93 nm. Fig. 2c clearly shows the moir$\acute{e}$  pattern as a periodic arrangement of bright and dark regions, with a wavelength of approximately 1.93 nm, confirming that it corresponds to the twist angle of 7.3$^\circ$. Based on these comprehensive characterizations, we conclude that the as-grown GSs consist of intertwined helical structures with a 7$^\circ$ in-plane twist, as depicted in the schematic illustrations in Fig. 2d.

To obtain a precise characterization of the out-of-plane structure of the GS and further confirm the presence of intertwined helical structures, cross-sectional TEM samples were prepared. These samples were extracted from the identical GS after performing CLSM observations (Fig. S6), TEM measurements (Fig. 2b), and magnetoresistance measurements (Fig. 1 and 3). The site-specific focused ion beam (FIB) lift-out technique was employed to prepare the GS cross-sectional samples. The process of FIB preparation for observing the cross-sectional structure of the GS is depicted in Supplementary Information Fig. S6.

\begin{figure*}[!htbp]
\includegraphics[width=0.8\textwidth,clip]{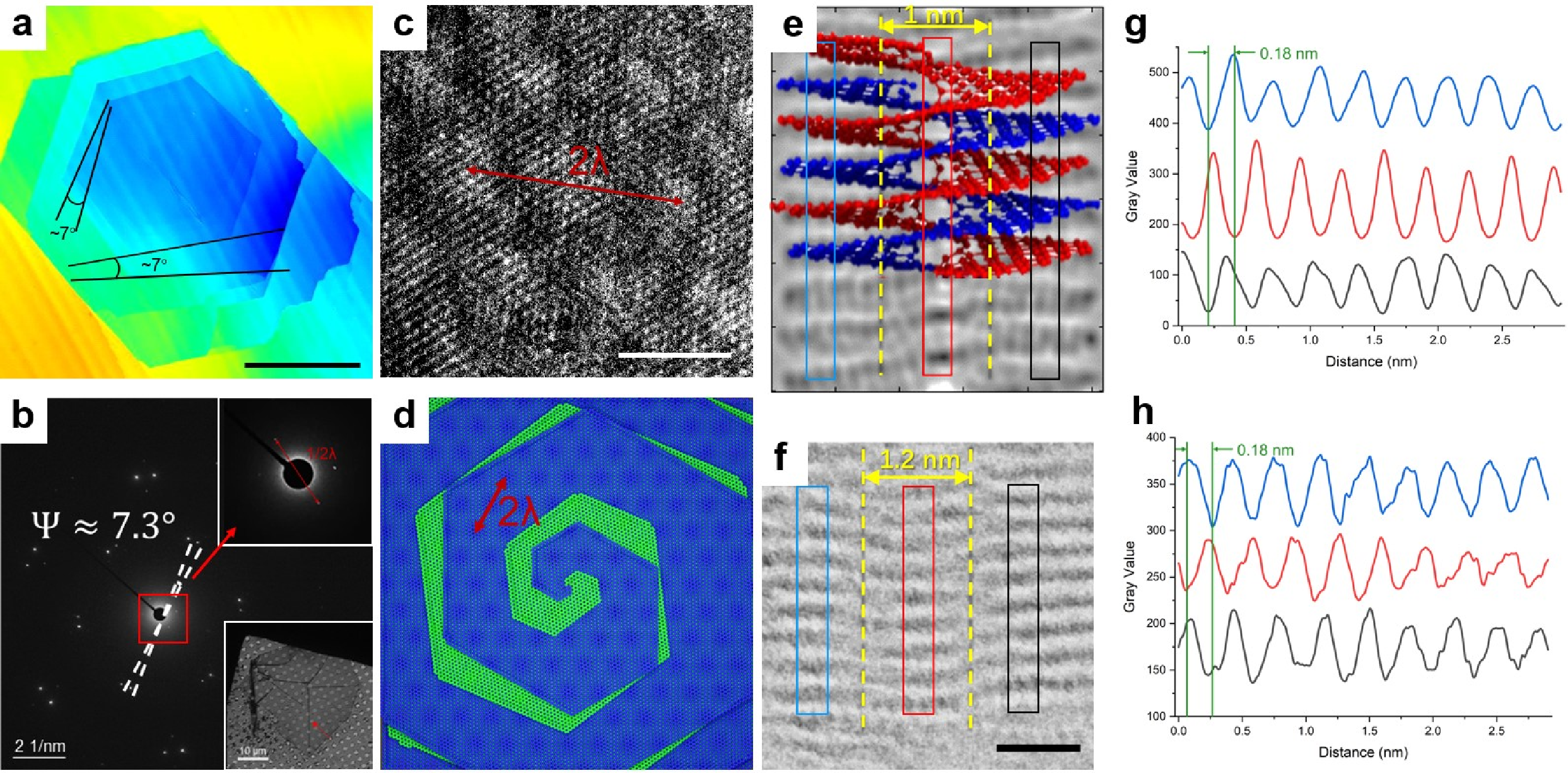}
\centering
\caption{(color online).  Structure analysis of the twisted graphene spiral. (a) The CLSM image with height information illustrates the central region of an individual GS; scale bar, 50 $\mu$m. (b) confirming an average twist angle of $\Psi \approx$ 7.3$^{\circ}$, which is consistent with the measurements obtained from AFM (shown in Fig. S8). The red rectangle in (b) highlights the presence of diffraction points corresponding to the moir$\acute{e}$ pattern, as further depicted in the top inset. The bottom inset shows a top-view TEM image of the GS transferred onto a TEM grid. (c) The moir$\acute{e}$ pattern as a periodic arrangement of bright (AA staking area) and dark (AB staking area) regions; scale bar, 2 nm. (d) Schematic diagram illustrating the structure of the GS. (e) Side view of the double helix GS model, and the corresponding simulated STEM image. (g) Contrast the line profile along the perpendicular plane across the colored strips in (e). A staggering series of graphene planes are signatures of double helical organization, as its simulated STEM a show. (f) Magnified STEM image of GS showing double helix feature. The contrast line profile analysis exhibits similar features as shown in the simulated image ((e) and (g)). TEM operating voltage 80 kV in order to minimize beam damage, scale bar, 1 nm.
}
\end{figure*}

The cross-sectional scanning transmission electron microscopy (STEM) observations (Fig. S7) reveal a double staggered series of graphene planes, suggesting the presence of a screw dislocation with a double helical structure. In order to gain a better understanding of the cross-sectional structure of the screw dislocation in the graphene multilayers, we constructed a structural model for the double helical structure which is further relaxed using molecular dynamics simulations. Then, we perform STEM intensity simulations based on the fully relaxed double helical structure using QSTEM code package \cite{RN143}. Fig. 2e presents a schematic model of the double helical structure of GS, accompanied by simulated STEM images derived from the fully relaxed structural model. Figs. 2e-h present a comparison between the configuration of the simulated double helical structure and the experimentally observed images (Figs. 2e, f). While the atomic structure cannot be completely determined from the cross-sectional STEM image, the stacking configuration and curvature closely resemble the expected double helical structure. This can be visualized as a result of cutting through the $sp^2$-bonded graphene layers perpendicularly, causing them to slip along the cutting line by two layers and reconnecting the bonds. Line profiles were plotted perpendicular to the lateral graphene layers, highlighting the distinctive configuration with double staggered graphene planes (Figs. 2g, h). This configuration involves a helical path traced around the linear defect (spiral core) as the atomic planes slip within the crystalline lattice. Indeed, a comparison of the line profile between the simulated double helical structure and the experimental observation reveals a remarkable similarity. This similarity is characterized by an integral-sliding pattern within the spiral core, where the atomic planes exhibit a shift equivalent to half a lattice of $sp^2$ carbon layers. The line profile analysis clearly shows a shift in the layered stacking configuration of approximately 0.18 nm (half the interlayer spacing of graphite) within the spiral core region (highlighted by the red stripes in Figs. 2e, f), compared to the surrounding areas (blue and black stripes). The remarkable consistency between the experimental line profile of the cross-sectional GS and the simulated image further confirms the proposed double helical structure.

\subsection{Quantum oscillation}

As well studied in previous works, a twist angle between two adjacent graphene layers can fundamentally change their electronic band structures, leading to topological flat bands which are responsible for various novel phenomena in electronic transport properties. The transport properties of the bulk twisted GS were investigated in a conventional four-terminal methods, and magnetic fields were applied parallel to the $c$-axis. The natural graphite (NG) was also studied as a comparison. A typical low temperature data for in-plane longitudinal resistivity ($\rho$) of twisted GS as a function of the magnetic field from $B$ = 0-9 T is shown in Fig. 3a. $\rho$ increases colossally with the magnetic field, and, at 9 T, it is almost 7 orders of magnitude larger than the zero-field value, which will be analyzed in next section.  In addition, Shubnikov-de Haas (SdH) oscillations are superimposed on the large magnetoresistance background. These oscillations, which start at a low magnetic field $B$ $\approx$ 0.1 T, can be better observed in the background-removed data plotted in Figure 2a and b. The background can be removed by subtracting a smoothed data curve. Fig. 3b shows oscillations under a smaller magnetic field region at different temperatures from 2 K to 16 K. Two series of oscillations can be distinguished. In the Fourier transformation of the $\Delta\rho$ versus (1/$B$) data shown in Fig. 3b inset, two frequencies are founded and assigned to electron pocket (4.67 T) and hole pocket (6.67 T). Peaks $e_{1,2}$, $h_{1,2}$ correspond to the 1$^{st}$ and 2$^{nd}$ harmonics of oscillations from electrons and holes. Thus, the area of fermi surface ($S_e=4.455\times10^{-4} \textrm{\AA}^2$ for electrons and $S_h=6.364\times10^{-4} \textrm{\AA}^2$ for holes) can be obtained throw the function $F=(\hbar/2\pi e)A_\textrm{F}$, where $F$ is the FFT frequency, $\hbar$ is reduced Planck’s constant, $e$ is elementary charge, $A_\textrm{F}$ is area of Fermi surface. Carrier mobility and concentration can reflect quasi-particle properties near the Fermi level, which are two important parameters of a material that can be derived from the Hall coefficient. 

\begin{figure*}[!htbp]
\includegraphics[width=0.8\textwidth,clip]{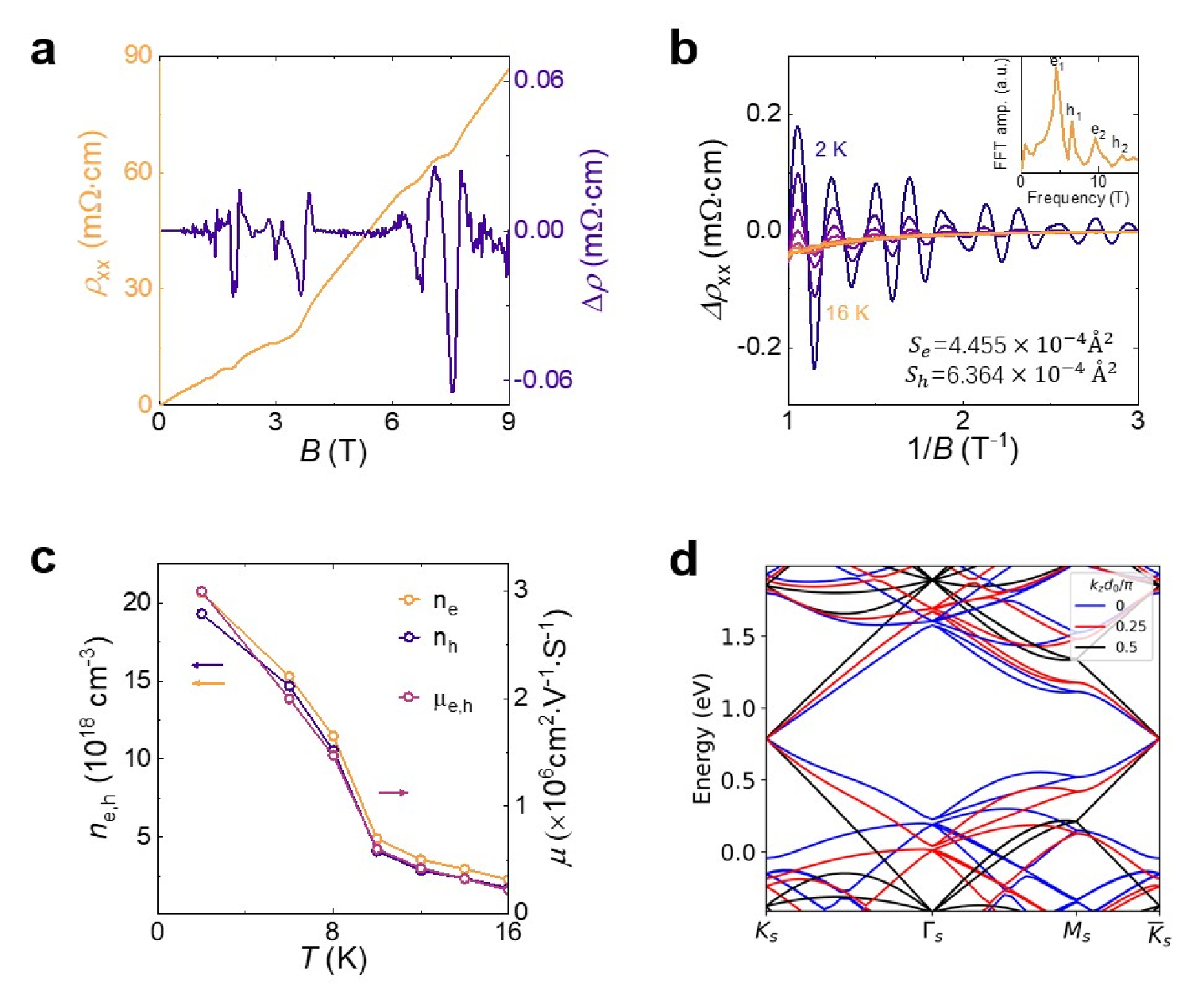}
\centering
\caption{(color online). Quantum oscillation. (a) Left axis: Resistivity $\rho_{xx}$ versus $B$ measured at $T$ = 2 K for GS. Right axis: Background removed data $\Delta \rho_{xx}$  showing quantum oscillations measured over different magnetic field regions. (b) Corresponding $\Delta \rho_{xx}$  data at different temperatures from 2 K to 16 K. The inset shows a FFT result at 2 K. Peaks $e_{1,2}$, $h_{1,2}$ correspond to the 1$^{st}$ and 2$^{nd}$ harmonics of oscillations from electrons and holes. Se and Sh indicate the fermi surface area of electrons and holes. (c) Temperature dependence of the carrier concentration (left ordinate) and the mobility (right ordinate). A relatively high carrier concentration and mobility can be observed at 2 K.  (d) The electronic band structures of GS in the ($k_x$, $k_y$) plane. The blue lines represent the electronic bands structures for $k_z d_0/\pi$=0. The red lines represent the electronic band structures for $k_z d_0/\pi$=0.25. The black lines represent the electronic band structures for $k_z d_0/\pi$=0.5.
}
\end{figure*}

Hall effect measurements have been performed in field sweep mode. The field dependence of the Hall resistivity ($\rho_{xy}$) exhibits a nonlinear behavior in low fields indicates the involvement of more than one type of charge carrier in the transport properties. The nonlinear Hall curve may be described by the two-carrier model \cite{RN234}: 
\begin{equation}
\rho_{xy}=\frac{1}{e} \frac{(n_h \mu_h^2-n_e \mu_e^2)+\mu_h^2 \mu_e^2 B^2 (n_h-n_e)}{(n_h \mu_h+n_e \mu_e)^2+\mu_h^2 \mu_e^2 B^2 (n_h-n_e)^2}B
\end{equation}
where $n_e$($n_h$) and $\mu_e(\mu_h)$ are the carrier density and mobility of electrons (holes), respectively. The mobilities of electron-type and hole-type are assumed to be equal at low temperatures, as both of them are quasi-particle excitations around the Dirac points which have similar scattering mechanisms. The fitting result can be seen in Figure 2c. The electron (hole) carrier concentration is found to be $n_e=2.07\times 10^{19}$ cm$^{-3}$ ($n_h=1.93\times 10^{19}$ cm$^{-3}$) at 2 K, with a net electron carrier concentration $\delta n_e$=1.4$\times$10$^{18}$ cm$^{-3}$. The corresponding mobility exhibits a high value $\mu_(e,h)=3 \times 10^6$  cm$^2$ V$^{-1}$ s$^{-1}$. This value is close to that of the Weyl semimetal NbP (5$\times$10$^6$  cm$^2$ V$^{-1}$ s$^{-1}$) \cite{RN23}.The quantum oscillation behavior may be qualitatively understood from the low-energy band structures of the bulk twisted intertwined GS. Away from the central dislocation core, the system can be viewed as numerous layers of TBG which are coupled via nearest neighbor interlayer hopping. As a result, such bulk moir$\acute{e}$ superlattice is bestowed an additional vertical wavevector ($k_z$) degree of freedom. When the twist angle is relatively large ($\geq 3^{\circ}$), the low-energy band structure consists of Dirac cones in the $k_x$-$k_y$ plane, with $k_z$-renormalized Fermi velocities, as shown in Fig. 2(d).  Neglecting the frame-rotation effects of the two graphene layers, it turns out that the low-energy states of the system have an approximate particle-hole symmetry at each vertical wavevector $k_z$. Consequently, the Dirac point at each $k_z$ is pinned to the same energy by virtue of this particle-hole symmetry. Including the frame-rotation effects would shift the energy of the Dirac point, leading to a $k_z$ dispersion of the form $4\gamma_0 cos^2 k_z d_0$ \cite{RN87}, with  $\gamma_0 \sim$ 3.5 meV, leading to a $k_z$ dispersion of Dirac points $\sim$7 meV.  This may lead to co-existence of electron and hole carriers even when the system is charge neutral. Moreover, other effects such as the variation of the interlayer distance (from 3.35 angstrom to 3.50 angstrom according to the TEM measurements) and remote-band renormalization effects may further enhance the $k_z$ dispersion of the Dirac points, which qualitatively explains the co-existing electron- and hole-type carriers in the system. As a two-band system with similar carrier concentration between electrons and holes, compensation mechanism can be a major factor. Considering the effect of topological Landau levels on MR, we measured $R$ ($T$) curves under different magnetic field.

\subsection{Anomalous metal-insulator transition}
Fig. 4a shows the temperature dependence of normalized $\rho$ of twisted GS and NG under zero magnetic field. The twisted GS shows a much larger residual resistivity ratio (RRR) [$\rho$(300 K)/$\rho$(2 K)] = 38, which is considerably larger than that of NG (RRR = 6). Generally, the less residual resistivity demonstrates high-quality of single crystals with less impurity scattering. However, the special lattice structure of the bulk twisted GS may also induce significant resistivity difference due to the dramatic changes in the band structures, as has been observed in twisted bilayer graphene \cite{RN37,RN42}.

\begin{figure*}[!htbp]
\includegraphics[width=0.8\textwidth,clip]{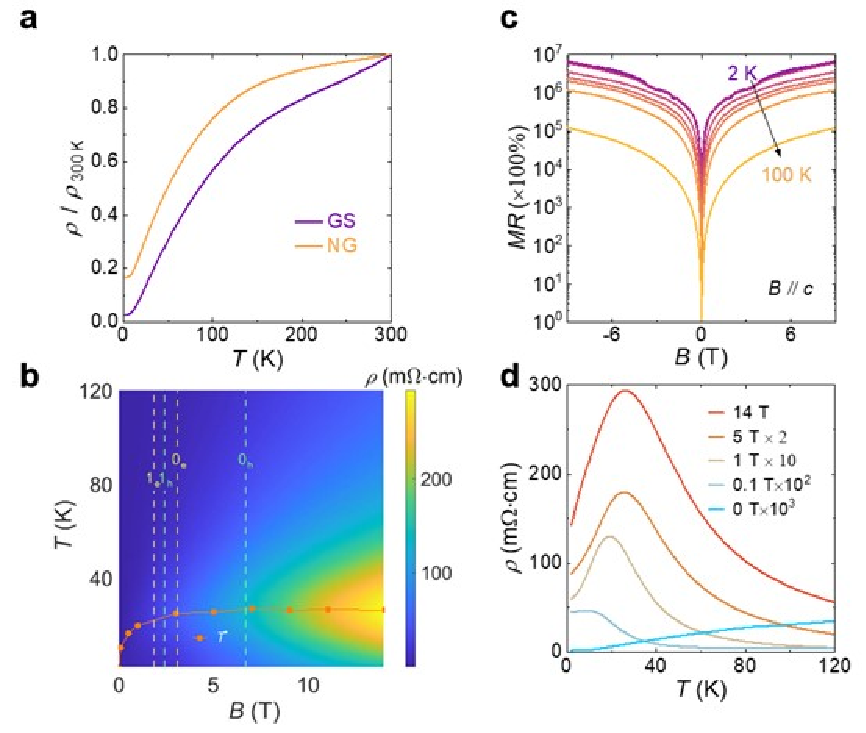}
\centering
\caption{(color online).Anomalous metal-insulator transition in twisted graphene spiral. (a) The normalized temperature dependent resistivity of GS and NG under zero field. (b) A color map of resistivity in the plane of the temperature and magnetic field. The orange scatterline indicates the magnetic field dependence of the anomalous transition temperature from metal to isolator. The dashed lines mark the filling factors of Landau levels from the electron and hole carriers. (c) MR of GS at different temperatures from 2 to 100 K. The highest MR of 7$\times$10$^6$\% is observed at 2 K under 9 T. (d) Line cuts of resistivity versus temperature for a range of magnetic field (shown are traces taken from 0 T to 14 T). An unconventional transition from metal to isolator appears when an finite external magnetic field has been applied.
}
\end{figure*}

A color map of resistivity in the plane of the temperature from 2 to 120 K and magnetic field from 0 to 14 T has been shown in Fig. 4b. An unconventional transition from metallic behavior (where resistivity $\rho$ increases with temperature) to insulator-like behavior (where resistivity $\rho$ decreases with temperature) appears when a finite external magnetic field has been applied. Transition temperature $T^\ast$ has been defined where $d\rho /dT$ = 0. The orange scatterline shows the field dependence of $T^\ast(B)$. As the magnetic field increases, the  $T^\ast$ rises rapidly and finally saturates to $\sim$26.9 K when the field is larger than 3 T. This is exactly the same magnetic field value above which the zeroth landau level starts to be filled, as marked by the rightmost yellow dashed line in Fig. 4b. Line cuts of resistivity versus temperature for several typical values of magnetic fields varying from 0 to 14 T can be seen in Fig. 4d. A monotonic temperature dependence where $\rho$  increases with temperature has been observed under zero magnetic field; and a nonzero  $T^\ast \sim$2 K emerges when the field is increased to 0.1 T, which is the same characteristic field amplitude for the onset of Landau level quantization. This implies that the anomalous metal-to-insulator transition at temperature  $T^\ast$ may be attributed to the correlation effects of the electrons in partially occupied Landau levels. Especially, in such bulk twisted GS system the linear Dirac dispersions and the quasi-2D Fermi surface would lead to weakly dispersive higher Landau levels and even dispersionless zeroth Landau level in such a 3D bulk system, which may significantly boost electron-electron interaction effects.

Line profiles of MRs under different temperatures are presented in Fig. 4c. This colossal magnetoresistance is concomitant with an anomalous metal-to-insulator transition with increasing temperature, which is reminiscent of isospin Pomeranchuck effect observed in magic-angle twisted bilayer graphene\cite{RN40,RN20}. A temperature dependence of MR under magnetic field ranging from 3 T to 14 T (shown in Fig. S3) exhibits an ultrahigh MR reaching 3.4$\times$10$^4$\%  under 14 T even at room temperature 300 K. Undisputedly, the topological dispresionless Landau level in such large-angle-twisted GS have enhanced the MR under a high magnetic field, although the specific mechanism still deserves further research.

\section{Conclusion}
In conclusion, our work has revealed a CMR and a remarkable metal-to-insulator transition in a bulk graphite superlattice arranged in a 3D TGS system. Through quantum oscillation measurements, we identified two distinct frequencies that correspond to separate charge carrier pockets: an electron pocket at 4.67 T and a hole pocket at 6.67 T. These findings are corroborated by Hall effect measurements. The simultaneous presence of both electron-type and hole-type carriers suggests a disruption in the particle-hole symmetries, possibly influenced by variations in the interlayer spacing and remote-band renormalization effects.

Upon applying an external magnetic field, we observed a notable transition from metallic to insulator-like behavior. The critical temperature, $T^\ast$, stabilizes around 26.9 K in fields above 3 T, indicating the onset of filling in the 3D zeroth Landau level. This behavior implies that the transition may stem from correlation effects among electrons in these partially filled 3D Landau levels. Accompanying this transition is a colossal magnetoresistance, measured at approximately 1.7$\times$10$^7$\%  at 2 K and 14 T, setting a new record. This significant increase in magnetoresistance is linked with the metal-to-insulator transition and is reminiscent of the isospin Pomeranchuck effect observed in magic-angle twisted bilayer graphene.

Our results establish the TGS as a potential new platform for studying CMR in 2D layered materials, alongside correlated and topological phenomena typically observed in moiré graphite superlattices.
\section{Methods}

\textbf{Structure analysis}.Transmission electron microscopy analysis. The Polymethyl Methacrylate (PMMA) solution was spin-coated onto the GS crystal at a speed of 600 rounds per minute (rpm) for 5 s, followed by spinning at a speed of 3000 rpm for 60 s and then drying at room temperature for 10 min. Following the drying process, the prepared sample was transferred onto a gold grid, and the PMMA protective layer was removed through a soaking procedure in acetone. Subsequently, Aberration-corrected transmission electron microscopy was carried out using a JEOL Grand Arm 300F, operated at 80 kV to minimize the risk of inducing knock-on damage to graphene. Image acquisition was performed with exposure time of 0.5 s on a OneView camera binned to $2k \times 2k$ pixel resolution. Post-processing of acquired images was undertaken for further analysis and enhancement of data quality.

\textbf{Theoretical analysis}. The structural relaxation is calculated utilizing Large-scale Atomic-Molecular Massively Parallel Simulation (LAMMPS)\cite{RN41} with long-rang bond-order potential for carbon (LCBOP) \cite{RN39}. A moir$\acute{e}$ supercell with periodic boundary condition applied in both the in-plane and out-of-plane directions, serves as the initial lattice structure. The twist angles between two layers, in accordance with the commensurate condition, are $\pm$7.34$^{\circ}$. The steepest descent algorithm is employed in the lattice relaxation calculations. The calculation of the band structure of twisted GS is performed utilizing the tight-binding (TB) model \cite{RN165} based on the full relaxed structure. To be specific, the hopping amplitude between two $p_z$ orbitals at different sites is expressed as:
\begin{equation}
t_{\text{\textbf{d}}} =V_{\sigma}(\frac{\textbf{d} \cdot \hat{\textbf{z}}}{d})+V_{\pi}[1-(\frac{\textbf{d} \cdot \hat{\textbf{z}}}{d})^2]
\end{equation}
where $V_{\sigma}=V_{\sigma}^0 e^{-(r-d_c)/\delta_0}$  and $V_{\pi}=V_{\pi}^0 e^{-(r-a_0)/\delta_0}$ $d = (d_x,d_y,d_z)$ is the displacement vector between two sites. $d_c$ = 3.44 $\text{\AA}$ is the interlayer distance. $a_0=a/\sqrt{3}$=1.42 $\text{\AA}$ is the distance between two nearest neighbor carbon atoms. $\delta_0$ = 0.184$a$. $V_\sigma^0$ = 0.48 eV and $V_\pi^0$= -2.7 eV.

\bigskip
\section{Acknowledgements}
This research was supported in part by the Ministry of Science and Technology (MOST) of China (Nos. 2022YFA1603903 and 2020YFA0309601), the National Natural Science Foundation of China (Grants Nos. 12174257, 12004251, 12104302, 12104303, 12304217), the Natural Science Foundation of Shanghai (Grant No. 20ZR1436100), the Science and Technology Commission of Shanghai Municipality (Grant No. 21JC1405100), the Shanghai Sailing Program (Grant No. 21YF1429200), the start-up funding from ShanghaiTech University, and Beijing National Laboratory for Condensed Matter Physics, the Interdisciplinary Program of Wuhan National High Magnetic Field Center (WHMFC202124).



\bibliography{1}

\end{document}